\documentclass[twocolumn,preprintnumbers,showpacs]{revtex4}
\usepackage{epsfig}
\usepackage{graphicx}
\begin{document}
%%%%%%%%%%%%%%%%%%%%%%%%%%%%%%%%%%%%

\author{}
\title{Generation of Superposition Spin States in an Atomic Ensemble}
\date{December 2, 2002}
\author{S. Massar}
\altaffiliation[Also at: ]{Ecole Polytechnique, C.P. 165, 
Universit\'e Libre de Bruxelles, 1050
Brussels, Belgium}
\affiliation{Service de Physique Th\'eorique, Universit\'e 
Libre de Bruxelles, C.P. 225,
Bvd. du Triomphe, 1050 Bruxelles, Belgium}
\author{E. S. Polzik}
\affiliation{QUANTOP - Danish National Research Foundation 
Centre for Quantum Optics,
Niels Bohr Institute, Copenhagen, DK-2100, Denmark}

\begin{abstract}
A method for generating a mesoscopic superposition state of the
collective spin variable of a gas of atoms is proposed. The state
consists of a superposition of the atomic spins pointing in two
slightly different directions. It is obtained by using off resonant
light to carry out Quantum Non Demolition Measurements of the
spins. The relevant experimental conditions, which require very dense
atomic samples, can be realized with presently available
techniques. Long-lived atomic superposition states may become useful as an
off-line resource for quantum computing with otherwise linear
operations.
\end{abstract}

\maketitle

Generation of superpositions of macroscopically different quantum
states have been attracting fundamental interest for a long time
\cite{S} since they illustrate the superposition principle in an
extreme setting. Recently, such states have also been proposed as an
off-line entanglement resource for a quantum computer, with which the
computer can work using only linear operations \cite{Milburn}.
Non-deterministically produced ``cat'' states of an ensemble of
particles serve in this proposal as an alternative to operations with
maximally entangled quantum states in other quantum computing schemes.

Superpositions involving a few microwave photons in a cavity have
been demonstrated in \cite{haroche}. Quantum superpositions have been
recently also demonstrated using currents in superconducting interference
devices \cite{SQUID}. There are proposals for realising such states in
different physical systems, see for instance \cite{proposals} for those
which involve Bose Einstein Condensates (BEC). Methods for generating
superpositions of two coherent states $|\alpha \rangle +|-\alpha \rangle $
of the electro-magnetic (e.m.) field have been suggested, but they require a
perfect measurement of the photon number \cite{optics}, which is impractical
at present.

In this paper, we propose a method for realizing a superposition state
of the collective spin variables of a gas of alkali atoms. Specifically the
state that would be realised is a coherent superposition of the atomic spins
pointing in two slightly different directions. Our method, in which the
state of the atomic spins 
is generated via an interaction with light pulses, is another
step within the growing field devoted to developing a quantum interface
between light and atomic ensembles.
Strong coupling between collective degrees of freedom of multi-atom
ensembles and multi-photon pulses of light can be achieved in free
space and has been recently used
to demonstrate squeezing of the collective spin
\cite{spinsqueezed,kuzmandel} and entanglement between two atomic
samples \cite{duan,Julsgaard}. This should be contrasted with the
strong coupling of individual photons to individual
atoms, which requires high finesse cavities \cite{haroche,cavity}.

The method we propose here for
generating a coherent superposition of the spins 
advances these
techniques beyond the continuous variables approximation. Indeed in
our proposal a strong atom-photon coupling has to be
realized with a small number of atoms. The relevant
experimental conditions, which require very dense atomic samples, seem
to be realisable   using presently available techniques.
The transfer of the atomic ensemble spin state onto the state of light is
also in principle possible \cite{Kuzmich,teleport}, and therefore
superpositions of coherent states of light could also be realised using our
method.

The collective spin variables for the ground state of an atomic ensemble $
{\hat{\vec{J}}}=\sum_{i}\hat{\vec{\j}}^{(i)},$ where $\hat{\vec{\j}}^{(i)}$ 
is the total angular
momentum of the $i$'th atom in the ground state, obey the usual commutation
relations $[\hat{J}_{x},\hat{J}_{y}]=i\hat{J}_{z}$. 
For simplicity we assume that each atom is
a spin $1/2$ system. We start with the coherent spin state with
maximal total angular 
momentum (i.e. all spins parallel) whereupon $
\hat J_{x}^{2}+\hat J_{y}^{2}+\hat J_{z}^{2}=
{\frac{N^{a}}{2}}({\frac{N^{a}}{2}}+1)$ where $
N_{a}$ is the number of atoms. We will suppose that the 
evolution of the collective spin is
limited to 
unitary transfirmations without dissipation. Under these conditions
the total length of the spin vector is conserved, i.e. it stays on the
surface of the same Bloch sphere. Throughout the paper the spin vector
stays polarized close to the $+z$ direction (close to $|\uparrow
_{z}\rangle $ 
state), but not necessarily exactly in this direction. We can then
define the rescaled variables 
$\hat x_A= \sqrt{2 \over N_a}\hat J_{x}$, $\hat p_A= \sqrt{2 \over
N_a} \hat J_{y}$, $\hat n_A={N_{A}\over 2}-\hat J_{z}$,
 which satisfy 
\begin{eqnarray}
\lbrack \hat x_A, \hat p_A] &=&i+O(\hat n_A/N_{A}) \ ,  \label{commut2} \\
{\frac{\hat x_A^{2}}{2}}+{\frac{\hat p_A^{2}}{2}} &=& \hat n_A+{\frac{1}{2}}%
+O(\hat n_A^{2}/N_{A})\ .  \label{number}
\end{eqnarray}%
Thus $\hat x_A$ and $\hat p_A$ are analogous to the position and momentum
operators, or to the quadrature operators of a single mode of the e.m.
field, and $\hat n_A$, the number of spins polarised in the $+z$ direction,
is analogous to the occupation number of a harmonic
oscillator, or to the number of photons in a mode of the e.m. field.
The new feature of the present model, as
compared to the earlier work, 
is that it goes beyond the harmonic oscillator approximation since the
deviation of $\hat J_{z}$ from $N_{a}/2$ is taken into account to first
order.

\begin{figure}[htb]
\centerline{\includegraphics[width=1\linewidth]{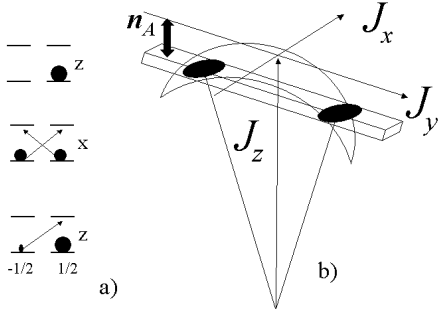}}
\caption{Generation of the superposition spin state. 
a) Atomic levels and interaction with light. Both ground and
excited states are spin $1/2$ states. The top diagram shows the initial
coherent spin state in $z$ representation. The medium diagram shows
the same state in $x$ representation along with the virtual
transitions used for generation of the spin squeezed state. The bottom
diagram shows the spin squeezed state in $z$ representation along with
the interaction used for the cat state generation. b) The spin squeezed
state generated by a QND measurement using the first pulse of light is
represented by the "banana" shape. The following QND measurement of
the number operator 
$\hat n_A$ corresponds to intersecting the "banana" with a plane of a
finite width. Thus the final state has two 
components lying at positive and negative values of $J_y$, i.e. it is a
superposition state.}
\label{Lfig2}
\end{figure}

The qualitative outline of the procedure for the ``cat'' state
generation goes as follows, see figure \ref{Lfig2}. As shown in
\cite{kuzmandel,Kuzmich,Julsgaard} a component of the collective 
spin operator can be measured in a QND way. 
A measurement of $\hat J_{x}$ to precision $\Delta 
J$ produces
a squeezed state of the atomic spin with uncertainties $\Delta
x_A^2=2\Delta J^2 /N_{a}=1/(2\xi^2) $ and $\Delta p_A^2=\xi^2/2 $
where $\xi $ denotes the degree of squeezing. Such a state has a
"banana" shape localized around the pole of the Bloch sphere.
The critical next step is to carry out a QND measurement of $\hat J_{z}$,
i.e. of the number operator $\hat n_A$, with the same precision $\Delta J$.
Such a measurement can be visualized as intersecting the "banana" with 
a plane parallel to the equatorial plane \ref{Lfig2}. The distance from the
plane to the pole depends on the result of the
QND measurement of $\hat n_A$. The "width" of the plane depends on the
precision $\Delta J$ with which $\hat n_A$ is measured. The resulting state
lies at the intersection of the plane and the initial squeezed
state. It is a superposition state.
Of central importance is that only ``weak'' QND measurements of the
atomic population, i.e. measurements of $\hat n_A$ 
which do not need to have the single atom accuracy, 
are required to produce the superpostion state.

Let us now describe in detail the QND measurements of the atomic spin. The
starting point is the effective interaction between the off resonant light
and the atomic ground state spin \cite{kuzmandel,Kuzmich,duan}. Atomic
levels and interactions
with light are shown in figure 1a). Let us suppose that initially the
photons are in the coherent state $|\psi_{P}\rangle=\int d x_P\exp
[-x_P^{2}/2]| x_P\rangle $ (hereafter we omit for simplicity
normalisation constants) and that the 
atoms are polarised in the $z$ direction in the
coherent spin state $ |0\rangle_{n_A} =\int dx_A\ \exp \left[
-x_A^{2}/2\right] 
|x_A\rangle$ (the level scheme on top in figure 1a).  For light propagating
along the $x$ axis the effective interaction Hamiltonian is $H_{eff}=\kappa
\sum_{i}|\uparrow _{x}\rangle _{i}\langle \uparrow _{x}|\hat a_{L}^{\dagger
}\hat a_{L}+|\downarrow _{x}\rangle _{i}\langle \downarrow _{x}|\hat
a_{R}^{\dagger 
}\hat a_{R}$ where $\hat a_{L}$, $\hat a_{R}$ 
are the annihilation operators for left and
right circular polarized light (the interaction is pictorially shown
in the middle level scheme 
in figure 1a). We introduce the Stokes operators $
\hat S_{x}=\hat a_{L}^{\dagger }\hat a_{L}-\hat a_{R}^{\dagger }\hat
a_{R}$, $\hat S_{y}=\hat a_{L}^{\dagger
}\hat a_{R}+\hat a_{R}^{\dagger }\hat a_{L}$, $\hat S_{z}=-i\hat 
a_{L}^{\dagger }\hat a_{R}+i\hat a_{R}^{\dagger
}\hat a_{L}$ and assume that $\hat S_{z}\simeq N_{p}$ 
is close to its maximal value
(where $N_{P}$ is the number of photons in the beam). Then we can define the
rescaled position-momentum like operators $\hat S_{x}=\hat 
x_P\sqrt{N_{p}/2}\ ,\
\hat S_{y}=\hat p_P/\sqrt{N_{P}/2}$ which obey $[\hat x_P,\hat p_P]=i$.
With this
notation the unitary evolution can be written (up to an unimportant overall
phase) as 
\begin{equation}
U=\exp \left( -ia\hat J_{x}\hat S_{x}\right) =\exp \left( -i\alpha
\hat x_A \hat  x_P\right)
\label{a}
\end{equation}
where $\alpha =a\sqrt{N_{a}N_{p}}/2$. 

In the first step a spin squeezed state 
is obtained as follows \cite{kuzmandel,Kuzmich,duan}
After the interaction with light the state is, in the $p_P$
basis, $U|\psi_{P}\rangle|0\rangle_{n_A}=\int dx_A dp_P \exp [-(\alpha
x_A-p_P)^{2}/2]\exp \left[
-x_A^{2}/2\right]|x_A\rangle |p_P\rangle$. Upon measuring $
\hat p_P$ a particular outcome $p_P$ is obtained.
The state of the atoms conditioned on the measurement outcome $p_P$ is
$$
\psi _{A}(p_P)=\int dx_A\ \exp [-(\alpha
x_A-p_P)^{2}/2]\exp \left[
-x_A^{2}/2\right]|x_A\rangle
 \ .  
$$
This is a squeezed state. By a rotation one can center the state
around 
$p_P=0$ whereupon one has
\begin{eqnarray}
\psi _{A}^{squeezed} &=&
 dx_A\ \exp [-(\alpha^2 + 1) x_A^{2}/2 ]|x_A\rangle
\nonumber\\
&=& \sum_{n_A} c(n_A)|n_A\rangle
\label{squ}
\end{eqnarray}
where
$c(n_A)=0$ if $n_A$ is odd, and equals
$c(n_A) = \left({\xi^2 -1\over 2 (\xi^2 +1)}\right)^{n_A/2}
{\sqrt{n_A!}\over
 ({n_A/2})!}$ when $n_A$ is even, with $\xi^2 = \alpha^2+1$.

The next step is generation of the cat state from the spin squeezed state 
by carrying out a QND measurement of $\hat J_z$. Towards this
end we send off resonant
right circular polarised light in a coherent state with the mean
number of photons $N_{p}$ 
 along the $z$ axis through the sample (the bottom level scheme in figure 1a). The temporal
evolution of the light and spins is then given by $U=\exp \left(
-ia\sum_{i}|\downarrow \rangle _{i}\langle \downarrow |\hat a_{R}^{\dagger
}\hat a_{R}\right)$. We define a translated creation operator by
$\hat a_{R}=\sqrt{N_{p}}+\hat a_{R}^{\prime }$ and the translated quadrature
operators $\hat x_{R}=(\hat a_{R}^{\prime }+\hat a_{R}^{\prime \dagger}
)/\sqrt{2}$, $\hat p_{R}=-i (\hat a_{R}^{\prime }-\hat a_{R}^{\prime \dagger
})/\sqrt{2}$ 
so that the unitary evolution takes the form
$U=\exp \left[ -ia \hat n_A \left( N_{p}+\sqrt{2N_{p}}\hat x_{R}\right)
\right]$ where we have dropped a term proportional to 
$\hat a_{R}^{\prime \dagger}\hat a_{R}^{\prime}$. The term proportional
to $N_{p}$ does not modify the state of the light, but it rotates the
atomic spin. This term can be taken into account at later stages of the
protocol and is omitted from now on.  
The part of the unitary transformation relevant for the
QND measurement is therefore 
$U=\exp \left[ -i\beta \hat n_A \hat x_{R}\right] $ where
$\beta =a \sqrt{2N_{p}}$. 

Initially light is in the coherent state $\psi _{R}=\int
dx_{R}\exp \left[ -x_{R}^{2}/2\right] |x_{R}\rangle $ 
and the spin squeezed state of atoms
 is given by eq. (\ref{squ}). After the interaction of light and
atoms, one measures the $p_{R}$
quadrature and finds an outcome $p_{R}$. The state of the atoms
conditional on this outcome is 
\begin{equation}
\psi_A (p_{R})= \sum_{n_A} c(n_A) 
\exp \left[ - (\beta n_A- p_{R})^{2}/{2}
\right] |n_A\rangle \ .
\label{cat11}\end{equation}
For suitable values of $\alpha, \beta, p_R$, this is a cat state. 
In order to show this one should  express the state in the
$x_A$  and $p_A$ basis. 
 This can be done using the matrix
elements $\langle p_A |n_A\rangle =\pi
^{-1/4}(2^{n_A} n_A!)^{-1/2}H_{n_A}(p_A)\exp [-p_A^{2}/2]$ where $
H_{j_{z}}(j_{y})$ are the Hermite polynomials. A numerical calculation
of $\psi_A (p_{R})$ in the $p_A$ basis is given in figure 2 (solid line).
To facilitate the analysis of the obtained state we derive approximate
analytical expressions for 
$\psi_A (p_{R})$ in the
$x_A$ and $p_A$ basis. First note that using the Stirling
expansion, the coefficients $c(n_A)$ can be approximated by
$c(n_A) \simeq \left({\xi^2 -1 \over  \xi^2 +1}\right)^{n_A/2}$ when
$n_A$ is even.
Inserting this expression into eq. (\ref{cat11}) we find that
\begin{eqnarray}
&\psi_A (p_{R})\simeq  \sum_{n_A\ even}  
\exp \left[ - \beta^2 (n_A- \mu(p_{R}))^{2}/{2}
\right] |n_A\rangle\ \  &
\label{cat12}\\
&\mbox{where } 
\mu(p_R)= {p_R \over \beta} + {1 \over 2 \beta^2} 
\ln \left({\xi^2 -1 \over  \xi^2 +1}\right)\simeq{p_R \over \beta}\ .&
\label{mumu}\end{eqnarray}
The distribution of number states is thus a Gaussian with the
mean $\mu(p_R)$ and the variance $1/\beta$. We now use the fact that 
the initial state is strongly squeezed in $x_A$, see
eq. (\ref{squ}). Thus we can use the approximation
 $n_A \simeq p_A^2 /2$ in eq. (\ref{cat12}). 
This implies that the wave
function in $p_A$ representation is centered around the two values 
$\pm \sqrt{2\mu(p_R)}$
with the variance $1/ (\sqrt{2\mu(p_R)} \beta)$. Hence the wave function is
\begin{eqnarray}
|\psi (p_{R})\rangle &=&\int dp_A\ \left( \exp \left[ -(p_A-\sqrt{2\mu }
)^{2}\beta ^{2}\mu \right] \right.  \nonumber  \label{psimux} \\
&&\left. +\exp \left[ -(p_A+\sqrt{2\mu })^{2}\beta ^{2}\mu \right]
\right) |p_A\rangle  \label{caty} \\
&=&\int dx_A\ \exp \left[ -{\frac{x_A^{2}}{4\beta ^{2}\mu }}\right] \cos 
\left[ x_A\sqrt{2\mu }\right] |x_A\rangle \quad \label{catx}
\end{eqnarray}
where the second equality if obtained from the first by taking the Fourier
transform, and the notations are simplified as $\mu =\mu (p_{R})$. 
Note that the Hermite polynomials with
even index are even functions of $p_A$ which explains the $+$ sign
between the two terms in eq. (\ref{caty}). As seen in figure 2,
eq. (\ref{caty}) is a good approximation to the exact wave function
(dotted line).

To verify that the cat state has indeed been obtained, the
measurements of spin components $\hat x_A$ and $\hat p_A$ 
should be carried out. The
results of the $\hat p_A$ measurement should be distributed according to
the 2 broad peaks described by eq. (\ref{caty}).  The condition for
observation of two distinct peaks 
is that the width of each of them should be less than the distance
between them.  The  
$ \hat x_A$ measurement should exhibit an
interference pattern due to the cosine in eq. (\ref{catx}). This
interference pattern is the proof that the two peaks found in the
$\hat p_A$ measurement are coherent, and therefore that a quantum
superposition between atomic spins pointing in two different
directions has been created. The condition for observation the
interference is that 
the period of the cosine should be 
less than the width of the Gaussian in (\ref{catx}). In fact both conditions
lead to the same inequality, $\mu \geq 1 / \beta$.
This condition means that by the QND measurement of the population 
one has removed the components with small $n_A$, 
i.e. the components near the origin $p_A = 0$, from the
squeezed state leaving a state with two peaks.

There is a second condition which is that the average population $\mu$
is limited by the 
degree of spin squeezing obtained in the first step of our protocol:
$\mu \leq \xi^2$. This 
condition has two origins. The first is that the probability of
finding values of $\mu$ which exceed $\xi^2$  is exponentially
small. The second is that when this condition is violated it becomes
impossible to measure the interference pattern in $x_A$ basis, and
thereby to prove that the state so obtained is indeed a coherent
superposition. 

Putting these two conditions together we obtain the condition
$\beta \xi^2 > 1 $ which relates the precision $1 /\xi$ with which one
can measure the $\hat x_A$ quadrature to the precision $1/ \beta$ with
which one can measure the number operator $\hat n_A$. Using the expressions
for $\beta$ and $\xi$ in terms of $a$, we can reexpress this
condition as
\begin{equation}
\xi ^{2}\geq N_{a}^{1/3}\ .  \label{cond1}
\end{equation}
Thus a superposition state can be produced if a high degree of
squeezing can be realised with a small number of atoms.
In what preceeds we have assumed that the quantum states are pure and
that there is no noise 
present. Below we consider decoherence of the atomic state
caused by the non-ideal QND measurements and describe a possible
experimental realization 
of the proposal.

The off resonant probe light used for the QND measurements will 
cause (weak) real excitations 
of atoms leading to an incoherent spontaneous process \cite{duan}. 
The significance of this process  can be assessed
in terms of accessible experimental parameters.
A first key parameter is the resonant optical depth of the atomic sample,
 $\kappa _{0}=\sigma N_{a}/A$ where $\sigma $ is the cross
section. The forward scattering amplitude of light is proportional to
$\frac{\gamma }{2}/({ i\Delta +\gamma /2)}$ where $\gamma $ is the
linewidth and $\Delta $ is the detuning with respect to
resonance. Hence the optical depth at detuning $ \Delta \gg \gamma $
is $\kappa _{\Delta }=\kappa _{0}\gamma ^{2}/4\Delta ^{2} $ and the
phase shift (or polarisation rotation) of light traversing
the sample is $\theta _{\Delta }=\kappa _{0}\gamma /2\Delta $.  The
parameter $a$ defined in eq. (\ref{a}) is the phase shift or
polarisation rotation per atom, hence equal to $a=\theta _{\Delta
}/N_{a}$.  The degree of squeezing that can be obtained is $\xi
=a\sqrt{N_{a}N_{p}}/2$.  Decoherence caused by the probe light can be described by the degree of
optical depumping, $\eta =\kappa _{\Delta
}N_{p}/N_{a}$ which can be interpreted as the probability that an
atom makes an incoherent transition. All these parameters can
be put together in the equation  $\xi ^{2}=\kappa _{0}\eta /4$
relating the degree of squeezing, the resonant optical depth and the
degree of optical depumping. 
The degree of optical
depumping must obey the condition $\eta \leq 1/\xi ^{2}$ otherwise the
coherence of the squeezed state or the superposition state will be
destroyed. This condition follows from the fact that a fraction $1/\xi
^{2}$ of the spins are pointing in the $-z$ direction and the other
spins are pointing in the $+z$ direction. Hence on average $
N_{A}/\xi ^{2}$ atoms can decohere before the coherence of the superposition
state, or of the parent squeezed state, is seriously affected. Thus the figure
of merit for generation of squeezing and entanglement is the resonant
optical depth of the sample which limits the degree of spin squeezing
to $\xi ^{2}\leq \frac{1 }{2}\sqrt{\kappa _{0}}.$ Using these
conditions and eq. (\ref{cond1}), one finds that the condition for
realising a superposition state can be written as
\begin{equation}
\kappa _{0}\geq 4N_{a}^{2/3}\ ,  \label{cond2}
\end{equation}%
that is a large optical depth must be achieved with a small number of
atoms.  This means that sending light through the sample along orthogonal
directions for the two steps of the protocol is not optimal. Rather it seems 
more practical to confine the atoms to a cylinder
of area $A$ and length $l$, with light sent along the long axis of the
cylinder, and to rotate the atomic spins between successive
measurements.  

The condition eq. (\ref{cond2}) can be achieved in, e.g., an alkali metal
BEC. Atomic density of the order of $10^{15}cm^{-3}$ for $4\cdot
10^{5}$ atoms has been reported in \cite{Zimmerman} for a cigar shaped
condensate with the optical density along the long axis on the order
of $10^{4}.$ This is marginally close to the condition
eq. (\ref{cond2}). \ The degree of squeezing required for cat state generation under the above
conditions is $\xi ^{2}\approx 50$. An additional improvement in the efficiency of the spin
measurement of a small number of atoms can be achieved by placing the
atomic sample in a low-finesse symmetric standing wave cavity with the
mirror transmission coefficients $T=1-R$. If the single pass
absorption probability and the single pass phase rotation angle
$\theta $ are much smaller than $T$, then the phase shift/polarisation
rotation of light exiting the cavity is $\theta ^{c}=2\theta /T$ and
the degree of optical depumping becomes $\eta ^{c}=2\eta /T$. Thus the
effect of the cavity can be summarized by introducing an effective cross section
$\sigma ^{c}=2\sigma /T$. The condition eq. (\ref{cond2}) for
realisation of a mesoscopic superposition state therefore becomes
$2\kappa _{0}/T>4N_{a}^{2/3}$. For a BEC sample with the density of $
10^{15}cm^{-3}$ placed in a low finess cavity with $T=5\%$, this
condition is met with $10^{3}$ atoms, which according to
eq. (\ref{cond1}) requires a feasible degree of spin squeezing $\xi
^{2}\approx 10$ for generation of the superposition state. 
Note that the life
time of the coherent superposition state is close to the life time of
the parent spin squeezed state, $\tau _{c}\xi ^{-2} $ \cite{wineland},
where $\tau _{c}$ is the coherence time of the atomic ground state,
which for a BEC is at least $100 m\sec $ \cite{BECtime}.

Finally let us mention some additional experimental conditions. We have
assumed above that the initial atomic spin state is perfectly polarized. In
practice the degree of spin polarisation obtained by optical pumping in a
vapor can reach $99\%$ \cite{Julsgaard}, which should be sufficient to
generate squeezing up to approximately $\xi ^{2}=100$. The degree of spin
polarization in a BEC can be even higher because only certain magnetic
states can be trapped. Another challenge associated with the cylindrical
configuration is that the atomic spins must be very precisely rotated
between each QND measurement (using either optical or radio frequency
pulses). It is easy to check that the precision of this rotation must
satisfy $\Delta \theta <1/(\xi ^{2}\sqrt{N_{a}})$ which for the free space
scenario with a BEC corresponds to the precision $\Delta \theta \simeq
3\times 10^{-5}$. For the low finesse cavity scenario a lower precision of $
\Delta \theta =1/300$ is necessary.

In summary we have proposed a method for generating a coherent superposition
of states in which the collective spin of an atomic ensemble points in two
slightly different directions. It appears to be realisable using 
present technology
by combining the methods used for QND measurements of collective atomic
variables with a BEC in a bad cavity. Preliminary experimental work
towards realisation of this proposal is under way.
This mesoscopic superposition state uses
the ground state atomic spins which means that it is relatively long lived.
It can therefore in principle 
be used as an offline non deterministic resource for
quantum logic as described in \cite{Milburn}. We will report on this
in a future publication and in particular we will show that 
the methods described here can be used  for purifying 
 the entanglement between two distant atomic ensembles.

We acknowledge financial support from the
Danish National Research Foundation, from the Communaut\'e
Fran\c{c}aise de Belgique under grant ARC 00/05-251, from the IUAP
programme of the Belgian government under grant V-18, from  
the EU through projects  EQUIP, RESQ, QUICOV and CAUAC.

\begin{figure}[htb]
\centerline{\includegraphics[width=0.7\linewidth]{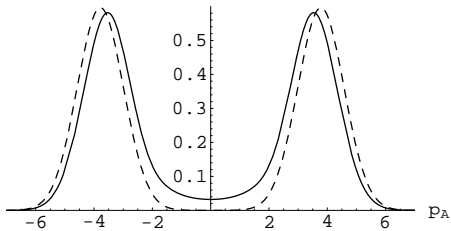}}
\caption{
The solid line shows the amplitude of the state $\psi_A(p_R)$, 
eq. (\ref{cat11}), 
in the $p_A$ basis for
parameters  $\xi^2=20$, $p_R/ \beta = 7$, $\beta = 1/3$. The dotted
line shows the approximation eq. (\ref{caty}) for the same values of the
parameters. 
}\label{fig2}
\end{figure}

\end{document}